\documentclass[twocolumn,showpacs,preprintnumbers,amsmath,amssymb,prb]{revtex4}
\usepackage{graphicx}% Include figure files
\usepackage{dcolumn}% Align table columns on decimal point
\usepackage{bm}% bold math

\begin{document}

\preprint{APS/123-QED}

\title{Lattice Dynamics of Metal-Organic Frameworks (MOFs): 
Neutron Inelastic Scattering and First-Principles Calculations}% Force line breaks with \\

\author{W. Zhou$^{1,2}$}
\author{T. Yildirim$^{1,2}$}%
\affiliation{%
$^{1}$NIST Center for Neutron Research, National Institute of Standards and
Technology, Gaithersburg, Maryland 20899,
USA\\$^{2}$Department of Materials Science and Engineering, University of
Pennsylvania, Philadelphia, Pennsylvania 19104, USA}%

\date{\today}% It is always \today, today,

\begin{abstract}
By combining neutron inelastic scattering (NIS) and first-principles 
calculations, we have investigated the lattice dynamics of MOF5. The 
structural stability of MOF5 was evaluated by calculating the three cubic 
elastic constants. We find that the shear modulus, $c_{44}$ = 1.16 GPA, is 
unusually small, while two other moduli are relatively large (i.e. 
$c_{11}$ = 29.42 GPa and $c_{12}$ = 12.56 GPa). We predict that MOF5 is very 
close to structural instability and may yield interesting new phases under 
high pressure and strain. The phonon dispersion curves and phonon density of 
states were directly calculated and our simulated NIS spectrum agrees very 
well with our experimental data. Several interesting phonon modes are 
discussed, including the softest twisting modes of the organic linker.
\end{abstract}

\maketitle

Metal-organic framework (MOF) 
compounds,\cite{Eddaoudi:2002}$^{,}$\cite{Yaghi:2003}$^{,}$\cite{Chae:2004}$^{,}$\cite{Ockwig:2005} 
which consist of metal-oxide clusters connected by organic linkers (see
Fig. 1), are a new class of nanoporous materials with very promising 
potential applications such as energy storage, gas separation, and template 
synthesis of nanoclusters and catalysts. There are an almost exponentially 
growing number of studies on MOF materials, mostly focusing on the 
optimization of the metal-oxide clusters and/or the organic linkers to 
improve the gas adsorption properties. Much less attention is paid to the 
fundamental properties of this interesting class of materials such as their 
structural stability and lattice dynamics. Understanding the stability and 
dynamical properties is clearly needed in order to optimize these materials 
for desired properties. For example, it was recently proposed that the MOF 
structure can encapsulate C$_{60}$ molecules and result in superconductivity 
upon doping\cite{Hamel:2005}. In such hybrid structures, it is important to 
know the phonon spectrum of the MOF lattice and its coupling to the 
electronic structure. Furthermore, a lack of knowledge concerning MOFs 
lattice dynamics presents a significant obstacle in the quantum dynamics 
study of gas molecules (H$_{2}$, CH$_{4}$ etc.) adsorbed on 
MOFs\cite{Rowsell:2005}$^{,}$\cite{Zhou:1}. 

\begin{figure}
\includegraphics[width=6cm]{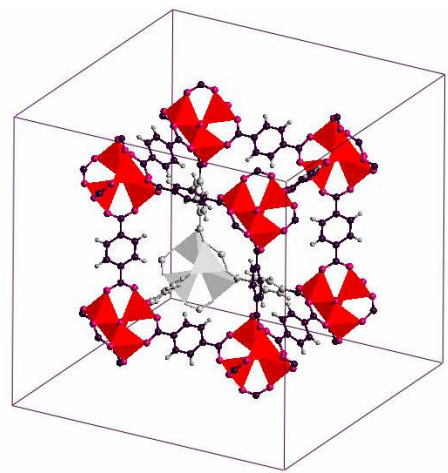}
\caption{
(Color online)
The crystal stucture of MOF5, which consists of BDC-linkers
connecting ZnO$_{4}$ clusters located
at the tetrahedral site of an fcc lattice.
For clarity, some of the atoms were not shown.
}
\label{fig1}
\end{figure}

Here we report a detailed study of the structural stability and lattice 
dynamics of MOF5 (the most widely studied MOF material) from combined 
neutron inelastic scattering (NIS) and first-principles calculations. The 
structure of MOF5 (Fig.1) is highly symmetric (space group \textit{Fm-3m}) and consists of 
ZnO$_{4}$ clusters at the tetrahedral site of an \textit{fcc} lattice, linked by 
1,4-benzenedicarboxylate (BDC) to form the three dimensional framework. 

\begin{figure}
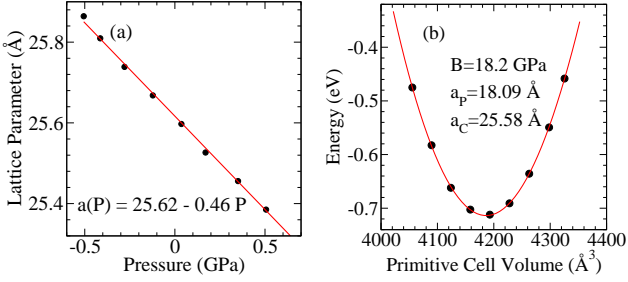

\includegraphics[width=4cm]{A_vs_P.eps} \;
\includegraphics[width=4cm]{E_vs_Vol.eps}
\caption{
(Color online)
Left: The cubic lattice parameter of MOF5 versus pressure, indicating linear
dependence near equilibrium structure. Right:
The total energy versus primitive cell volume (dots) and the
fit by Murnaghan equation of state (solid line). 
}
\label{fig2}
\end{figure}

Our first-principles calculations were performed within the plane-wave 
implementation of the local density approximation to density functional 
theory in the PWscf package\cite{Baroni:1}. We used Vanderbilt-type 
ultrasoft potentials with Perdew-Zunger exchange correlation. A cutoff 
energy of 408 eV and a 1$\times $1$\times $1 k-point mesh were found to be 
enough for the total energy to converge within 0.5meV/atom.

We first describe the dependence of the structure on the cell volume (i.e. 
external pressure). Figure 2 shows how the lattice parameter and the total 
energy of the cell changes with external pressure and the total volume, 
respectively. We optimized the atomic positions for each volume and did not 
find any structural instability over the pressure range studied. From -0.5 
to 0.5 GPa, we find a linear pressure dependence of the lattice constant 
with a slope of 0.46 {\AA}/GPa. Fitting the energy vs. volume curve to the 
Murnaghan equation of state\cite{Murnaghan:1944} yields a bulk modulus of 
18.2 GPa. The minimum energy corresponds to lattice constant $a_{C}$ = 25.58 
{\AA}, which is in excellent agreement with the experimental 
value\cite{Yildirim:2005} of $a$ = 25.91 {\AA}. The optimized atomic positions 
were also found to agree well with the experimental values.

\begin{figure}
\includegraphics[width=8cm]{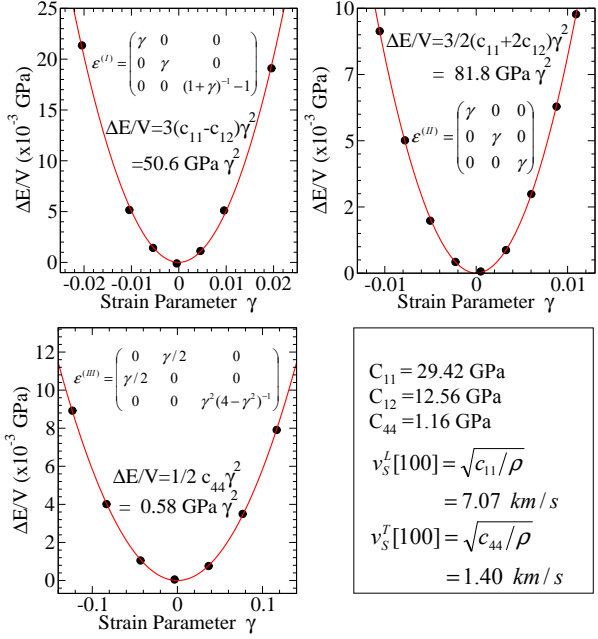}
\caption{
(Color online)
The three cubic elastic constants of MOF5. The dots are
the actual calculations and the solid lines are the quadratic
fit. The deformation matrices for each distortion are also
shown. The right-bottom panel summarizes the calculated
elastic moduli and the sound-velocities. Here $\rho $ is the mass density of MOF5 (0.59 g/cm$^{3})$.
}
\label{fig3}
\end{figure}

A better insight into the structural stability can be gained from the 
elastic constants, which we calculated following the standard scheme for a 
cubic crystal\cite{Mehl:1994,Beckstein:2001}, starting from the fully optimized MOF5 
structure. Three types of strains were applied to the conventional fcc cell 
of MOF5 (see Fig.~3). 
Strain $I$ is a stretch along the z axis; strain \textit{II} is a monoclinic 
shear about the $z$ axis. Both strain $I$ and \textit{II} are volume-conserving. Strain \textit{III} 
simply changes the lattice constant $a$. For all three strains, there is only 
one variable, \textit{$\gamma $}, in the strain tensor, reflecting the magnitude of the 
strain. All calculations were done on the strained primitive cell, which is 
structurally equivalent to the strained conventional cell. Atomic positions 
were fully optimized and $E$ vs. \textit{$\gamma $} data were obtained. \textit{E($\gamma )$/V} was then fit to a 
polynomial in \textit{$\gamma $}, where $V$ is the unit cell volume. 
%The fitted coefficients of 
%\textit{$\gamma $}$^{2}$ corresponds to 3($c_{11} \quad - \quad c_{12})$, $c_{44}$/2 and 3($c_{11}$ + 
%2$c_{12})$/2 for the three types of strains, 
%respectively.\cite{Beckstein:2001} Note that strain \textit{III} alone can also give 
%bulk modulus by fitting the $E(V)$ data to a traditional Murnaghan equation of 
%state\cite{Murnaghan:1944}, as was done earlier. In Fig. 3, we show all the 
%calculated results along with the data fit. 
The results are summarized in Fig.~3.
As expected, the calculated elastic constants 
satisfy the stability criteria (i.e. $c_{11}  >  c_{12}$, $c_{11}$ + 
2$c_{12} >$ 0, and $c_{44}  >$ 0 for a cubic crystal). However, the shear 
modulus $c_{44}$ is very small (1.16 GPa), indicating that MOF5 is very close 
to structural instability. Besides, $c_{11}  -  c_{44}$ is significantly 
larger than zero and according to the Cauchy relation\cite{Born:1985}, this 
indicates a significant deviation from a central intermolecular potential. 
The small $c_{44}$ also suggests that MOF5 could collapse into a potentially 
useful structure under certain shear-stress. For examples, the pristine MOF5 
structure has a large pore volume but the pores are too open to hold 
hydrogen molecules at high temperatures; a collapsed MOF5 structure could 
have finer cavities connected by smaller channels, which are in principle 
better for H$_{2}$ storage.

After having shown that the MOF5 structure is stable with a very small shear 
modulus, we now discuss the lattice dynamics properties at the zone center 
and along high-symmetry directions in the Brillouin zone. The phonon density 
of states (DOS) and dispersion curves were calculated using the supercell 
method with finite difference.\cite{Yildirim:2000} The primitive cell was 
used and the full dynamical matrix was obtained from a total of 26 
symmetry-independent atomic displacements (0.03 {\AA}). 

The primitive cell of MOF5 contains two formula units 
(Zn$_{4}$O$_{13}$-(C$_{8}$H$_{4})_{3})$ giving rise to a total of 318 
phonon branches. The phonon modes at $\Gamma $ are classified as

$\Gamma $ (q=0) = 9$A_{1g}$(R) + 3$A_{1u}$ + 3$A_{2g}$ + 9$A_{2u}$ + 12$E_{u}$ + 
12$E_{g}$(R) + 17$T_{2u}$ + 24$T_{2g}$(R) + 24$T_{1u}$(IR) + 17$T_{1g}$

where R and IR correspond to Raman- and infrared-active, respectively. The 
crystal symmetry implies 105 Raman- and 72 IR-active modes. In Table I, we 
list the calculated energies at $\Gamma $. We hope that our calculations 
will initiate more experimental work such as Raman/IR measurements to 
confirm the gamma phonon energies that we calculated here. In the absent of 
such high-resolution data, we have performed neutron inelastic scattering 
measurements of the phonon DOS and compared it with our calculations.

\begin{table}[htbp]
\begin{center}
\caption{List of phonon symmetries and calculated energies (in meV) of MOF5 at the 
$\Gamma $ point of the primitive cell.}
\begin{tabular}{|ll|ll|ll|ll|ll|}
\hline\hline
$A_{2g}$& 
2.6& 
$A_{2u}$& 
23.7& 
3$T_{1g}$& 
74.3& 
$A_{1u}$& 
121.0& 
2$E_{g}$& 
171.0 \\
3$T_{1g}$& 
3.1& 
$A_{1g}$& 
28.7& 
3$T_{2g}$& 
74.6& 
3$T_{2u}$& 
121.0& 
3$T_{2g}$& 
171.3 \\
2$E_{g}$& 
3.5& 
3$T_{1u}$& 
31.7& 
3$T_{1g}$& 
75.4& 
2$E_{u}$& 
121.1& 
$A_{1g}$& 
171.9 \\
3$T_{2u}$& 
3.7& 
3$T_{2u}$& 
32.5& 
3$T_{2g}$& 
75.8& 
2$E_{u}$& 
124.8& 
3$T_{1g}$& 
183.0 \\
3$T_{2u}$& 
5.3& 
3$T_{1g}$& 
33.3& 
2$E_{g}$& 
81.9& 
3$T_{1u}$& 
124.9& 
2$E_{u}$& 
183.5 \\
3$T_{1u}$& 
6.8& 
$A_{2u}$& 
33.7& 
3$T_{2g}$& 
82.1& 
$A_{2u}$& 
124.9& 
3$T_{1u}$& 
183.5 \\
2$E_{u}$& 
8.0& 
3$T_{1u}$& 
34.2& 
3$T_{1g}$& 
82.1& 
3$T_{2u}$& 
131.7& 
$A_{2u}$& 
183.5 \\
3$T_{1g}$& 
9.5& 
3$T_{2g}$& 
34.3& 
3$T_{2g}$& 
82.2& 
3$T_{1u}$& 
131.7& 
3$T_{2g}$& 
183.6 \\
$A_{1u}$& 
10.1& 
2$E_{u}$& 
34.6& 
$A_{1g}$& 
82.7& 
2$E_{g}$& 
137.1& 
2$E_{g}$& 
190.6 \\
$A_{2g}$& 
10.8& 
3$T_{2g}$& 
43.0& 
3$T_{2u}$& 
88.8& 
3$T_{2g}$& 
137.2& 
3$T_{2g}$& 
190.7 \\
3$T_{2g}$& 
10.9& 
3$T_{1g}$& 
43.5& 
3$T_{1u}$& 
89.2& 
$A_{1g}$& 
137.2& 
$A_{1g}$& 
191.1 \\
3$T_{1g}$& 
11.0& 
$A_{1u}$& 
49.0& 
2$E_{u}$& 
95.7& 
2$E_{g}$& 
140.4& 
3$T_{2u}$& 
191.6 \\
3$T_{2g}$& 
13.5& 
3$T_{2u}$& 
49.1& 
3$T_{1u}$& 
96.3& 
2$E_{u}$& 
140.6& 
3$T_{1g}$& 
193.5 \\
3$T_{2u}$& 
13.6& 
2$E_{u}$& 
49.1& 
$A_{2u}$& 
97.5& 
3$T_{2g}$& 
140.6& 
3$T_{1u}$& 
196.2 \\
3$T_{1u}$& 
14.2& 
3$T_{2g}$& 
54.5& 
3$T_{1g}$& 
98.3& 
3$T_{1u}$& 
140.6& 
3$T_{2g}$& 
197.5 \\
2$E_{g}$& 
15.2& 
$A_{1g}$& 
55.3& 
3$T_{2g}$& 
98.5& 
$A_{2u}$& 
140.8& 
2$E_{u}$& 
381.6 \\
3$T_{1g}$& 
15.7& 
2$E_{g}$& 
55.9& 
2$E_{g}$& 
102.3& 
$A_{1g}$& 
141.2& 
3$T_{1u}$& 
381.6 \\
3$T_{1u}$& 
16.1& 
3$T_{2u}$& 
56.1& 
3$T_{2g}$& 
102.6& 
3$T_{1g}$& 
154.3& 
$A_{2u}$& 
381.6 \\
3$T_{2u}$& 
17.4& 
3$T_{1u}$& 
56.1& 
$A_{2g}$& 
103.0& 
3$T_{2g}$& 
154.3& 
3$T_{1g}$& 
381.7 \\
2$E_{u}$& 
18.5& 
3$T_{1u}$& 
57.8& 
$A_{1g}$& 
103.2& 
2$E_{u}$& 
167.9& 
3$T_{2g}$& 
381.7 \\
3$T_{2g}$& 
18.5& 
3$T_{2g}$& 
59.0& 
3$T_{1g}$& 
103.2& 
3$T_{1u}$& 
168.1& 
3$T_{2u}$& 
383.0 \\
3$T_{1g}$& 
18.9& 
$A_{2u}$& 
68.2& 
2$E_{g}$& 
103.3& 
$A_{2u}$& 
169.0& 
3$T_{1u}$& 
383.0 \\
3$T_{2u}$& 
19.7& 
2$E_{u}$& 
68.6& 
3$T_{2u}$& 
107.7& 
3$T_{2u}$& 
170.1& 
2$E_{g}$& 
383.2 \\
3$T_{2g}$& 
20.6& 
3$T_{1u}$& 
68.6& 
3$T_{1u}$& 
107.8& 
3$T_{1u}$& 
170.2& 
3$T_{2g}$& 
383.2 \\
2$E_{g}$& 
21.5& 
3$T_{2u}$& 
72.5& 
3$T_{1g}$& 
119.5& 
3$T_{2u}$& 
171.0& 
$A_{1g}$& 
383.3 \\
3$T_{1u}$& 
22.0& 
3$T_{1u}$& 
73.1& 
3$T_{2g}$& 
119.6& 
3$T_{1u}$& 
171.0& 
& 
 \\
\hline\hline
\end{tabular}
\label{tab1}
\end{center}
\end{table}

\begin{figure}
\includegraphics[width=7cm]{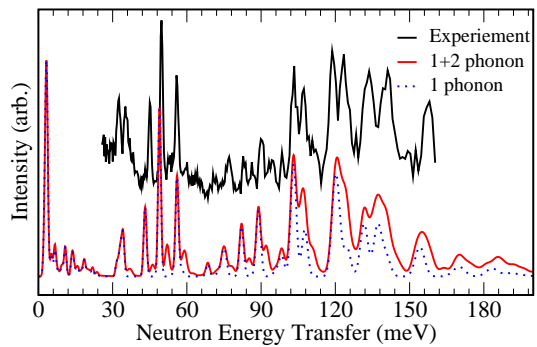} 
\caption{
(Color online)
Measured NIS spectrum of MOF5 at 4~K (top curve) 
along with calculated 
spectrum (bottom curves). 
The calculated 1 (dotted line)  and 1+2 phonon contributions are shown.
}
\label{fig4}
\end{figure}

\begin{figure}
\includegraphics[width=7cm]{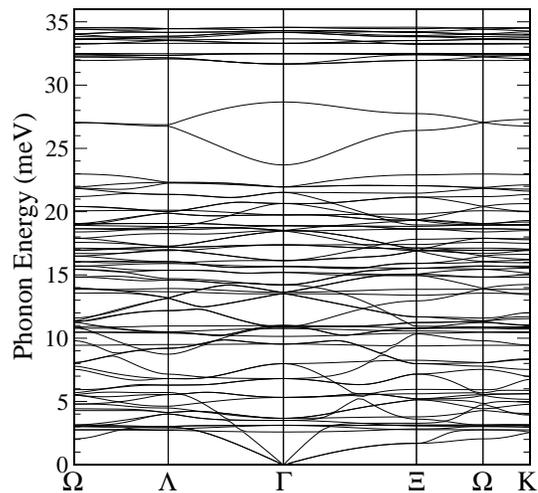}
\caption{
Calculated phonon dispersion curves along high-symmetry 
directions in the Brillouin zone for MOF5. For clarity, only low energy 
(0-40 meV) portion is shown. 
}
\label{fig5}
\end{figure}

The experimental NIS spectrum shown in Fig.4 was collected on a powder 
sample of MOF5 using the Filter Analyzer Neutron 
Spectrometer\cite{Udovic:2004} (FANS) under conditions that provide 
full-width-at-half-maximum energy resolutions of 2-4.5{\%} of the incident 
energy over the range probed. To compare the NIS data with theory, the NIS 
spectrum was computed for a 10$\times $10$\times $10 q-point grid within the 
incoherent approximation\cite{Yildirim:2000}$^{,}$\cite{Squires:1996}, 
with instrumental resolution taken into account. As shown in Fig. 4, the 
agreement between calculation and experiment is excellent. In contrast to 
Raman and IR spectroscopy, neutron spectroscopy is not subject to any 
selection rules, thus all modes falling in the energy range covered by our 
measurement ($\sim $20 to 170 meV) were detected. Within this $E$ range, the 
NIS spectrum is dominated by one-phonon processes. The two-phonon 
contribution coming from the combination of one-phonon processes is not 
significant. Also note that the NIS spectrum is dominated by hydrogen 
displacements because of the relatively large H neutron scattering cross 
section.

Figure 5 shows the dispersion curves of the low-energy modes along 
high-symmetry directions in the Brillouin zone. The high energy modes are 
quite dispersionless and thus not shown. It is interesting to point out that 
some of the modes have significant dispersion, such as those between 25-30 
meV. This is surprising, considering the rigidity of the units (i.e., the 
ZnO$_{4}$ cluster and the BDC linker). We find that the dispersion comes 
from the points where these two rigid units are connected. This is also the 
reason for the very low shear modulus $c_{44}$. Another interesting result is 
the very low energy modes near 3 meV, which are quite flat. Some of these 
phonon modes will be discussed in detail below. Up to now, no 
phonon-dispersion measurement on MOF5 has been reported, partly due to the 
difficulty of growing a large single crystal needed for the experiment. We 
hope that in the near future, some of the phonon dispersions that we predict 
here can be experimentally confirmed.

We next discuss several interesting modes revealed by the phonon 
calculations. Animations of these phonon modes are available in 
Ref.\cite{ncnr:1} for direct visualization. The lowest energy mode 
($A_{2g}$, 2.59 meV) is singly degenerate, with all C$_{6}$H$_{4}$ linkers 
twisting around the crystal axis direction. This mode is basically 
dispersionless (see Fig. 5) which is due to negligible interaction between 
two adjacent benzene linkers along the crystal axis. The second lowest 
energy mode ($T_{1g}$, 3.12 meV, 3-fold degenerate) can be described as two 
of three C$_{6}$H$_{4}$ linkers twisting around the two corresponding 
crystal axes. These two softest twisting modes imply that the bonds between 
the organic linkers and the metal clusters are the most unstable part of the 
MOF structure. The $\sim $0.53 meV energy difference between the two modes 
suggests a coupling of the twisting motions of the nearest-neighbor benzene 
linkers.

\begin{figure}
\includegraphics[width=7cm]{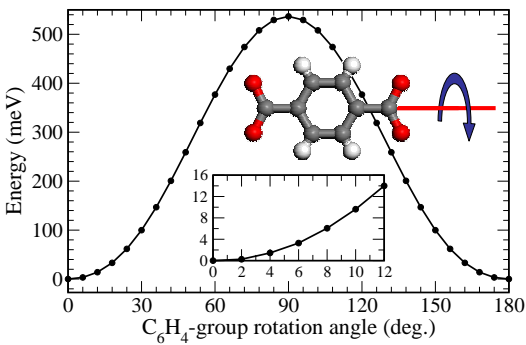}
\caption{
(Color online)
The variation of the total energy as one of the
 C$_{6}$H$_{4 }$ groups (part of the BDC organic 
linker of MOF5) is twisted around the crystal axis direction. 
Note that the initial twisting 
does not cost much energy (see inset) although the 
energy barrier for the 180$^{o}$ flipping (two-site jumping) of the 
C$_{6}$H$_{4 }$ group is rather large ($\sim $537 meV).
}
\label{fig6}
\end{figure}

Since the energy of the twisting mode is very low, one may wonder if the 
benzene ring is free to rotate at high temperatures. This could have 
important consequences on the hydrogen absorption dynamics in MOF5. If the 
linker is rotated 90$^{o}$, it could block the channels connecting the cubic 
cavities in MOF5 available for hydrogen storage. Motivated by these 
arguments, we have investigated the dynamics of the linker using 
quasielastic neutron scattering measurements\cite{Bee:1988}. However, we 
did not observe any free rotation of the linker at high temperature nor any 
evidence of the $\pi $-rotational jumps of the linker. These findings later 
became clear when we looked at the potential curve as the C$_{6}$H$_{4}$ 
linker is rotated. We performed total energy calculations as a function of 
the rotational angle of the benzene plane. For simplicity, we only twisted 
one of the six benzene rings in the primitive cell. The results are shown in 
Fig.6. The overall barrier is surprisingly large ($\sim $537 meV), 
consistent with the quasielastic neutron scattering results. However, 
twisting for a few degrees only costs an energy of several meV (see the 
inset of Fig. 6), in agreement with the phonon modes being very soft. The 
third lowest energy mode (3.50 meV, $T_{2u}$, 3-fold degenerate) involves all 
C$_{6}$H$_{4}$ linkers translating along crystal axis $a$, $b$, and $c$. All other 
modes with E $<$ 23 meV are various combinations of the motion of the 
C$_{6}$H$_{4}$ linker as a whole and the vibration of the ZnO$_{4}$ 
clusters. Two modes at slightly higher energy (23.69 meV, $A_{2u}$ and 28.67 
meV, $A_{1g})$ are the breathing modes of the ZnO$_{4}$ cluster, with the 
clusters vibrating in the unit cell out-of-phase and in-phase, respectively. 
Most of other modes falling in the energy range of our NIS measurement are 
one of the twisting, stretching, breathing, bending and wagging motions of 
the benzene rings. They can be roughly divided into two groups. With E $<$ 
$\sim $70 meV, the phonon modes are more dominated by the C displacements of 
the C$_{6}$H$_{4}$ linkers. The modes in the higher energy region are more 
dominated by H displacements.

The highest energy modes (380-385 meV, not covered by our NIS data) are 
totally dominated by the C-H covalent bond and all are associated with 
various C-H stretching modes of the C$_{6}$H$_{4}$ group.

In summary, we presented a detailed lattice dynamics study of MOF5 combining 
first-principles calculations and neutron inelastic scattering measurements. 
Our results indicated that MOFs are an interesting class of materials in 
terms of structural stability and dynamical properties. We identified the 
modes associated with the twisting of the linker to be the softest modes in 
the structure and therefore expected to be populated first with increasing 
temperature, eventually leading to structural instability. This is also 
consistent with the very small shear modulus c$_{44}$ that we have 
calculated. Despite the very low energy of the linker twisting modes, we 
found a very large energy barrier for the full rotation of the linker. We 
hope that our calculations will initiate further experiments to confirm our 
predictions. In particular, our results suggest that MOF5 is very close to 
structural instability, possibly leading to an interesting structural 
transformation under high pressure.

The authors thank Drs. T. J. Udovic and M. R. Hartman for technical help in 
the NIS data collection. We also acknowledge partial support from DOE under 
BES grant DE-FG02-98ER45701.


\begin{thebibliography}{11}
\bibitem{Eddaoudi:2002} M. Eddaoudi et al., Science \textbf{295}, 469 (2002).
\bibitem{Yaghi:2003} O. M. Yaghi et al., Nature \textbf{423}, 705 (2003).
\bibitem{Chae:2004} H. K. Chae et al., Nature \textbf{427}, 523 (2004)
\bibitem{Ockwig:2005} N. Ockwig, O. D. Friedrichs, M. O'Keeffe, O. M. Yaghi, Acc. Chem. Res. \textbf{38}, 176 (2005)
\bibitem{Hamel:2005} S. Hamel, V. Timoshevskii, and Michel Cote, Phys. Rev. Lett. \textbf{95}, 146403 (2005)
\bibitem{Rowsell:2005} J. Rowsell, J. Eckert, OM Yaghi. J. Am. Chem. Soc. \textbf{127}, 14904 (2005)
\bibitem{Zhou:1} W. Zhou and T. Yildirim, in preparation
\bibitem{Baroni:1} S. Baroni, A. Dal Corso, S. de Gironcoli, and P. Giannozzi, http://www.pwscf.org.
\bibitem{Murnaghan:1944} F. D. Murnaghan, Proc. Natl. Acad. Sci. U.S.A. \textbf{30}, 244 (1944).
\bibitem{Yildirim:2005} T. Yildirim and M. R. Hartman, Phys. Rev. Lett. \textbf{94}, 175501 (2005).
\bibitem{Mehl:1994} M. J. Mehl, B. M. Klein, and D. A. Papaconstantopoulos, "First-Principles Calculation of Elastic Properties," Intermetallic Compounds: Principles and Applications, ed. J. H. Westbrook and R. L. Fleischer (London: John Wiley {\&} Sons Ltd., 1994).
\bibitem{Beckstein:2001} O. Beckstein, J. E. Klepeis, G. L. W. Hart, and O. Pankratov, Phys. Rev. B \textbf{63}, 134112 (2001).
\bibitem{Born:1985} M. Born and K. Huang, in Dynamical Theory of Crystal Lattices (Oxford University Press, Oxford,1985), p. 136--140.
\bibitem{Yildirim:2000} T. Yildirim, Chem. Phys. \textbf{261}, 205 (2000).
\bibitem{Udovic:2004} T. J. Udovic, D. A. Neumann, J. Le\~{a}o, and C. M. Brown, Nuclear Instruments and Methods in Phys. Research A \textbf{517}, 189 (2004).
\bibitem{Squires:1996} G. L. Squires, Introduction to the theory of thermal neutron scattering (Dover, New York, 1996).
\bibitem{ncnr:1} http://www.ncnr.nist.gov/staff/taner/h2
\bibitem{Bee:1988} M. Bee, Quasielastic Neutron Scattering (Hilger, Bristol, 1988).
\end{thebibliography}
\end{document}